\documentclass[pra,twocolumn,showpacs,longbibliography,notitlepage]{revtex4}
\usepackage[cp1251]{inputenc}
\usepackage{hyperref}
\usepackage{natbib}
\usepackage{amsmath}
\usepackage{amssymb}

\usepackage{bm}
\usepackage{color}
\usepackage{epsfig}
\begin{document}

 \renewcommand {\Im}{\mathop\mathrm{Im}\nolimits}
  \renewcommand {\Re}{\mathop\mathrm{Re}\nolimits}
    \newcommand {\Tr}{\mathop\mathrm{Tr}\nolimits}
\newcommand {\rmi}{{\rm i}}
\newcommand {\rmd}{{\rm d}}
\newcommand {\sign}{\mathop{\mathrm{sign}}\nolimits}
\newcommand {\e}{{\rm e}}
\renewcommand {\phi}{\varphi}
\renewcommand {\epsilon}{\varepsilon}
\newcommand {\eps}{\varepsilon}
\newcommand{\ei}{\eps_{\rm Si}}
\newcommand{\ee}{\eps_{\rm eff}}
\newcommand{\eout}{\eps_{\rm out}}
\newcommand{\eo}{\eps_{\rm SiO_{2}}}
\newcommand{\nix}[1]{}
\title{Local field corrections to the spontaneous emission in   arrays of Si nanocrystals}
 \author{Alexander N. Poddubny}
 \affiliation{Ioffe Institute of the Russian Academy of Sciences, St Petersburg 194021, Russia}
 \email{poddubny@coherent.ioffe.ru}
\begin{abstract}
We present a theory of the local field corrections to the spontaneous emission rate for the array of silicon nanocrystals in silicon dioxide. An analytical result for the Purcell factor is obtained. We demonstrate that the local-field corrections are sensitive to the volume fill factor of the nanocrystals in the sample and are suppressed for large values of the fill factor. The local-field corrections and the photonic density of states are shown to be described by two different effective permittivities: the harmonic mean between the nanocrystal and the matrix permittivities and the Maxwell-Garnett permittivity.
\end{abstract}
\date{\today}
\pacs{42.70.Qs, 42.50.-p}
\maketitle

\section{Introduction}
Modification of the spontaneous emission rate in the electromagnetic environment  is well known since the seminal work of Purcell~\cite{Purcell}. This effect has been studied for various systems, including microcavities~\cite{kavbamalas}, photonic crystals, and metamaterials~\cite{review2013}. Here, we focus on the example of Si nanocrystals in SiO$_{2}$ matrix~\cite{Priolo2014}. The Purcell factor, defined as the ratio of the radiative decay rate of a single spherical nanocrystal in SiO$_{2}$ to that in vacuum, reads \cite{Yablonovitch88,Delerue2004}:
\begin{equation}
f_{\rm purc,0}\equiv\frac{\tau_{0}}{\tau_{\rm rad}}=\sqrt{\eps_{\rm SiO_{2}}}F(\eo),\quad F(\eps)=\left(\frac{3\eps}{\eps_{\rm Si}+2\eps}\right)^{2}\:,\label{eq:rad1}
\end{equation}
where $\tau_{0}$ is the radiative lifetime in vacuum (calculated neglecting the field screening), 
$\eps_{\rm SiO_{2}}$ and $\eps_{\rm Si}$ are the dielectric constants of the silicon dioxide matrix and silicon, respectively. The $\sqrt{\eps_{\rm SiO_{2}}}$ factor in Eq.~\eqref{eq:rad1} is determined by the photonic density of states in the matrix. The second factor $F(\eo)$ describes the local field corrections. It is equal to the squared ratio of the electric field inside the (spherical) nanocrystal to the field outside the nanocrystal and can be determined from the solution of a pure electrostatic problem~\cite{landau08}. Substituting the values $\eps_{\rm Si}\approx 12$ and $\eps_{\rm SiO_{2}}\approx 2$, we obtain $F(\eo)\approx 0.14$, i.e. the field inside the nanocrystal is strongly screened~\cite{Delerue2004}.

\begin{figure}
\centering\includegraphics[width=0.4\textwidth]{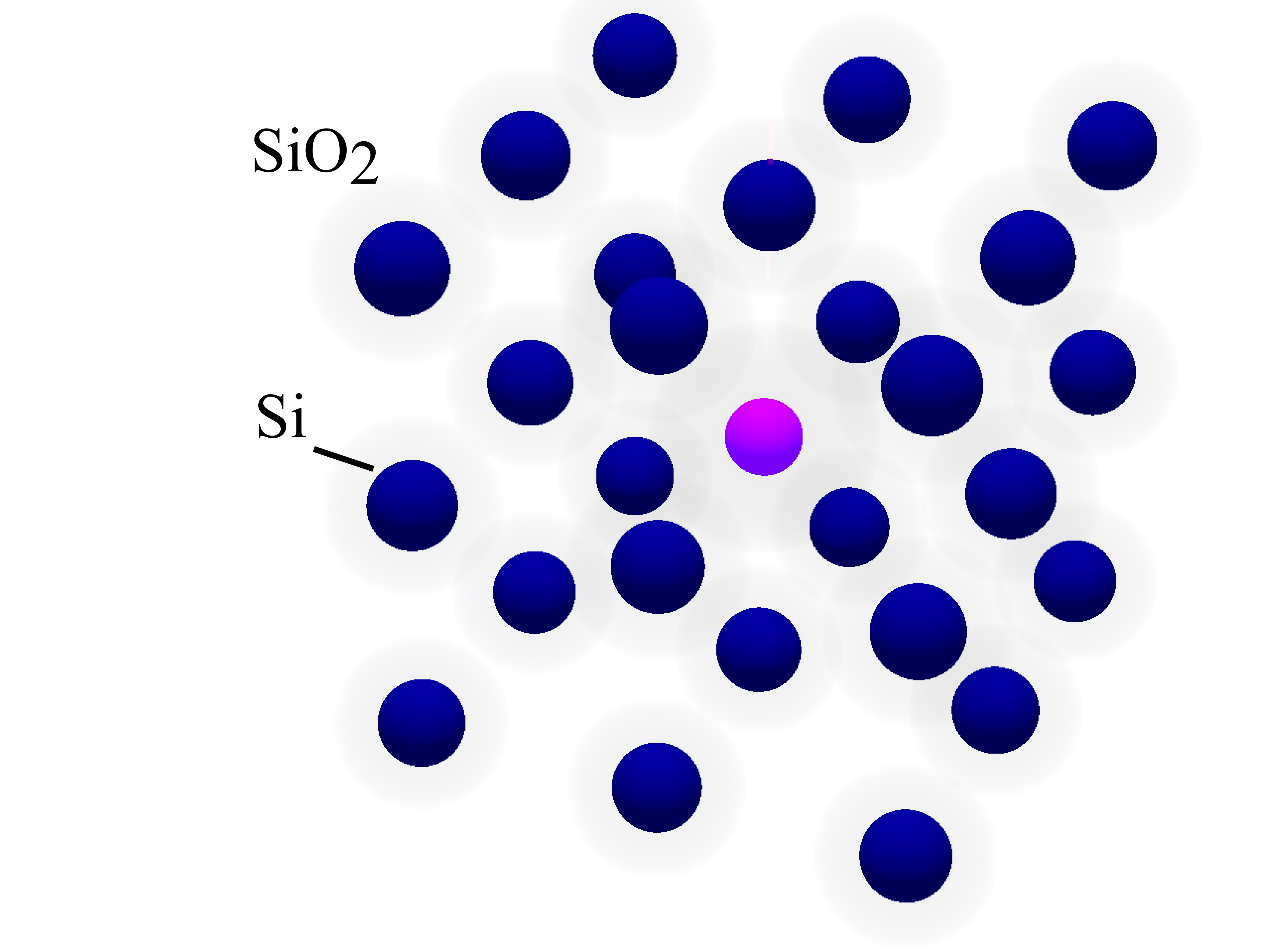}
\caption{Schematic illustration of the emission in the cubic array of Si nanocrystals}\label{fig:geometry}
\end{figure}

Equation~\eqref{eq:rad1} has been obtained for the case of single nanocrystal. In practice, however, one often considers  relatively dense samples with the volume fractions of Si exceeding $f=0.1$~\cite{Timmerman2011,we2014}. Since the permittivity of Si is quite large, the effective dielectric constant of such samples can considerably exceed that of empty SiO$_{2}$. As a result, one can expect strong modification of the radiative decay rate Eq.~\eqref{eq:rad1}: both the density of photonic states in the sample and the local field factor should be sensitive to the volume fraction of Si $v$. To the best of our knowledge, this problem has not been examined in the literature. Existing microscopic electromagnetic studies were limited to the case of single emitter and/or different geometries~\cite{Schuurmans,busch2000,Fussel2004,Rahmani2002,poddubny2012cross}. In particular,  in the previous works \cite{Rahmani2002,poddubny2012cross} the local field corrections   have been taken into account, however, the geometry of the problem was quite different. Instead of the emitting nanocrystal,  an external interstitial emitter has been introduced into the array of scatterers. In the current manuscript we address the problem of the local field corrections to the spontaneous emission in nanocrystal arrays (see Fig.~\ref{fig:geometry}) when the emitter is at the lattice site and the local field corrections are influenced by its neighbors in the lattice.

Solution of this problem is essential for the reliable measurement of the radiative emission rate of the nanocrystals.
Typically, the recombination of electron-hole pairs is contributed by both the spontaneous emission and the nonradiative recombination channels. Measurement of the decay kinetics of the photoluminescence only does not allow one to distinguish between these two processes. One of the established 
\cite{snoeks1995,blum2012,Fuji} approaches to separate the effects is to modify the radiative decay rate by placing the emitters close to the dielectric or metallic mirror. Assuming that the nonradiative decay stays the same, one can extract the radiative decay rate  from the dependence of the total decay rate on the distance from the mirror. However, the validity of this technique for the dense nanocrystal samples is not yet clear.

The rest of the paper is organized as follows. In Sec.~\ref{sec:Model} we present the calculation of the spontaneous emission rate for the cubic lattice of nanocrystals in the discrete dipole approximation. Sec.~\ref{sec:Discussion} is devoted to the discussion of the results, and Sec.~\ref{sec:Summary} is reserved for the summary.

\section{Coupled dipoles model}\label{sec:Model}
\subsection{Single nanocrystal}
We start with re-deriving the classical result Eq.~\eqref{eq:rad1} within the framework of the discrete dipole model and a semiclassical approximation. While the spontaneous emission {\it per se} is a quantum process, its modification in the medium can be studied semiclassically~\cite{ginzburg1983,Welsch2006}.
We describe the nanocrystal by the {\it bare} dipole polarizability
\begin{equation}\label{eq:alpha}
\alpha_{\rm bare, res}=\alpha_{\rm bare}+\frac{d^{2}}{\hbar(\omega_{0}-\omega)},\quad \alpha_{\rm bare}=V_{\rm NC}\frac{\ei-\eo}{4\pi}\:,
\end{equation}
where the first term accounts for the difference between the background dielectric constants of Si and SiO$_{2}$ with $V_{\rm NC}$ being the nanocrystal volume. The second term describes the  polarization due to the excitonic state with the resonance frequency $\omega$, the parameter $d$ is an effective dipole moment of the transition. Our goal is to determine the complex radiative correction to the frequency $\omega_{0}$, arising due to the interaction with light. To this end, we solve the coupled system of the equation for the total electric field $\bm E$ and the material equation for the dipole moment of the nanocrystal $\bm p$:
\begin{align}\label{eq:EP}
\bm E(\bm r)&=G(\bm r)\bm p+\bm E_{0}(\bm r)\\
\bm p&=\alpha_{\rm bare,res}\bm E(0)\nonumber\:,
\end{align}
where $\bm E_{0}$ is the incident field,
\begin{equation}\label{eq:G}
G(\bm r)=\frac{1}{\eo}(q^{2}+\nabla\otimes\nabla)\frac{\e^{\rmi qr}}{r}
\end{equation}
is the tensor Green function, $q=\omega\sqrt{\eo}/c$ is the light wave vector and $\nabla\otimes\nabla$ denotes the tensor $\partial^{2}/\partial x_{i}\partial x_{j}$.
Evaluating the first of Eqs.~\eqref{eq:EP} at $\bm r=0$  and $\bm E_{0}=0$ and combining it with the second one we obtain the equation for the eigenfrequencies of the system:
\begin{equation}
{\det}\left(\frac{1}{\alpha_{\rm bare, res}}-G(0)\right)=0\:. \label{eq:sec1}
\end{equation}
It should be noted that the real part of the Green function Eq.~\eqref{eq:G} diverges at the origin. To overcome this problem one has to take into account the finite spread of the emitter. This can be carried out either by replacing the local material relation $\bm p=\alpha_{\rm bare}\bm E(0)$ by the nonlocal one~\cite{Ivchenko2005,Goupalov2003}, or, alternatively, by keeping the local material relation but regularizing the Green function at the origin \cite{lagendijk_review}. Both approaches are equivalent for subwavelength nanocrystals when the dipole approximation holds. Here, we resort to the second approach, which means that $G$ at $\bm r=0$ is replaced by the regularized value
\begin{equation}
 G(0)=\frac{2\rmi q^{3}}{3\eo}-\frac{4\pi}{3V_{\rm NC}\eo}\:.\label{eq:G0}
\end{equation}
Substituting Eq.~\eqref{eq:G0} and Eq.~\eqref{eq:alpha} into Eq.~\eqref{eq:sec1} we obtain the closed-form equation for the complex eigenfrequency $\omega$. We use the weak-coupling approximation when the Green function can be evaluated at the resonant frequency $\omega_{0}$. For the nanocrystal with spherical symmetry the solution is three-fold degenerate.  The imaginary part of the eigenfrequency $\omega$ in the lowest order in the retardation parameter $q_{0}^{3}V$ reads
\begin{equation}
-2\Im \omega\equiv\frac1{\tau_{\rm rad}}=\frac{4\omega_{0}^{3}d^{2}}{3\hbar c^{3}}\sqrt{\eps_{\rm out}}\left(\frac{3\eps_{\rm out}}{\eps_{\rm in}+2\eps_{\rm out}}\right)^{2}\label{eq:Iomega}\:.
\end{equation}
This is equivalent to the known answer Eq.~\eqref{eq:rad1} with the standard expression for the radiative lifetime in vacuum \cite{Welsch2006} $1/\tau_{0}=4\omega_{0}^{3}d^{2}/(3c^{3}\hbar)$. 
The complex polarizability of the nanocrystal with and without excitons, renormalized by the interaction with light and describing the response to the incident field $\bm E_{0}$, is given by \cite{lagendijk_review}
\begin{equation}
\alpha_{\rm res}=\frac{\alpha_{\rm bare,res}}{1-\alpha_{\rm bare,res}G(0)},\quad
\alpha=\frac{\alpha_{\rm bare}}{1-\alpha_{\rm bare}G(0)}\:.
\end{equation}
When the radiative correction term $\propto q^{3}$ in Eq.~\eqref{eq:G0} for $G(0)$ is neglected, the expression for $\alpha$ reduces to the well-known electrostatic result~\cite{landau08} for the dipole polarizability of the single sphere
\begin{equation}
\alpha=\frac{3V_{\rm NC}\eo}{4\pi}\frac{\ei-\eo}{\ei+2\eo}\:.\label{eq:alphas}
\end{equation}
Re-derivation of Eq.~\eqref{eq:Iomega} and Eq.~\eqref{eq:alphas} justifies the validity of our approach and provides the route to generalize the technique for the array of nanocrystals.
\subsection{Array of nanocrystals}

 We consider the periodic cubic array with the period $a$, forming the lattice $\bm r_{j}$ (see Fig.~\ref{fig:geometry}). For such system, one can obtain closed-form analytical results in the long-wavelength approximation $c/\omega\gg a$~\cite{Purcell1973,sprik1996,belov2005,poddubny2012cross,gorlach2014}. 
The coupled-dipole equations describing the problem read
\begin{equation}
\bm p_{j}=\alpha_{j}\sum\limits_{j'\ne j}G(\bm r_{j}-\bm r_{j'})\bm p_{j'}\label{eq:coupling}
\end{equation}
with
\begin{equation}
\alpha_{j}=\begin{cases}
\alpha_{\rm res},& j=1\:,\\
\alpha,& j\ne 1\:.
\end{cases}
\end{equation}
Here,  we  take into account the resonant polarization term only in one of the nanocrystals, with $j=1$ and $\bm r_{j}=0$.  This corresponds to the situation of strong inhomogeneous broadening of the excitonic resonance.
The system Eq.~\eqref{eq:coupling} can be equivalently rewritten as
\begin{equation}
\frac1{\alpha}\bm p_{j}=\sum\limits_{j'\ne j}G(\bm r_{j}-\bm r_{j'})\bm p_{j'}+\delta_{j,1}\left(\frac{1}{\alpha}-\frac1{\alpha_{\rm res}}\right)\bm p_{1}\:.\label{eq:coupling2}
\end{equation}
In order to solve the system Eq.~\eqref{eq:coupling} the dipole momenta can be expanded over the Bloch eigenmodes
\begin{equation}
\bm p_{j}=\sum\limits_{\bm k}\bm p_{\bm k}\e^{{\rm i}\bm k\bm r_{j}}\:,\label{eq:pj}
\end{equation}
with $\sum_{\bm k}\equiv a^{3} \int {\rm d}^{3}k/(2\pi)^{3}$ and the integration being performed over the first Brillouin zone.
For the Bloch modes $\bm p_{\bm k}$ Eqs.~\eqref{eq:coupling2} become independent,
\begin{equation}
\frac1{\alpha}\bm p_{\bm k}=C_{\bm k}\bm p_{\bm k}+\left(\frac{1}{\alpha}-\frac1{\alpha_{\rm res}}\right)\bm p_{1}\label{eq:coupling3}
\end{equation}
with 
\begin{equation}
C_{\bm k}=\sum\limits_{j\ne 1}\e^{-\rmi \bm k\bm r_{j}}G(\bm r_{j})
\end{equation}
being the interaction constant~\cite{belov2005}. 
Solving Eq.~\eqref{eq:coupling3} for $\bm p_{\bm k}$,  and substituting the result back to Eq.~\eqref{eq:pj}  we obtain the following equation for the eigenfrequency $\omega$:
\begin{equation}
{\rm det}\left[\frac{\hbar(\omega-\omega_{0})}{d^{2}}-\frac{1}{\alpha_{\rm bare}}+\frac{\alpha}{\alpha_{\rm bare}^{2}}\sum\limits_{\bm k}\frac1{1-\alpha C_{\bm k}}\right]=0
\end{equation}
 The second term in this equation is real. Hence, the Purcell factor, i.e. the normalized spontaneous emission rate, is determined only by the third term. This term is proportional to the identity matrix due to the cubic symmetry of the problem. The final result can be presented as
\begin{equation}
f_{\rm purc}=\frac{3}{2(\omega/c)^{3}}\Im \frac{1}{\alpha_{\rm bare}^{2}}\sum\limits_{\bm k}\left[\frac1{1/\alpha- C_{\bm k}-\rmi 0}\right]_{zz}\:,\label{eq:fgen}
\end{equation}
where $q_{0}\equiv q(\omega_{0})$ and the interaction constant is to be evaluated at the frequency $\omega_{0}$\:.

It is instructive to analyze Eq.~\eqref{eq:fgen} in more detail. First, if the interaction constant is neglected, the result reduces to $f_{\rm purc}=3\Im \alpha/[2(\omega/c)^{3}\alpha_{\rm bare}^{2}]$. This expression is equal to $\sqrt{\eo}F(\eo)$ for $(\omega/c)^{3}V\ll 1$, in agreement with the result Eq.~\eqref{eq:rad1} for a single nanocrystal. Second, for the periodic structure the quantity $1/\alpha- C_{\bm k}$ is real~\cite{belov2005}. Hence, the imaginary part of Eq.~\eqref{eq:fgen} appears only due to the poles determined by the dispersion equation $1/\alpha- C_{\bm k,zz}=0$.
This is designated by the symbol $\rmi 0$ in the denominator of Eq.~\eqref{eq:fgen}: the integral is to be calculated by adding the negligibly small quantity $-\rmi \delta$ in the denominator and then taking the limit $\delta \to 0$.
 Physically, this means that the energy is carried away from the source only by the photonic eigenmodes of the structure, that satisfy the dispersion equation.

Now we proceed to the analytical evaluation of Eq.~\eqref{eq:fgen} in the long-wavelength limit $c/\omega_{0}\gg a$.
The wave vectors of the photonic modes responsible for the spontaneous emission are then much smaller than $1/a$. Hence, the interaction constant can be  presented in the Maxwell-Garnett approximation \cite{belov2005} as
\begin{equation}
C_{\bm k}=-\frac{2{\rm i} q_{0}^{3}}{3\eo}+\frac{4\pi}{3V_{0}\eo}+\frac{4\pi}{V_{0}\eo}\frac{q_{0}^{2}-\bm k\otimes\bm k}{k^{2}-q_{0}^{2}}\:,
\end{equation}
where $V_{0}=a^{3}$ is the unit cell size.
Before performing the integration we introduce the effective dielectric constant in the Maxwell-Garnett approximation \cite{Lagendijk1997}
\begin{equation}
\eps_{\rm eff}=\eo\left(1+\frac{4\pi\alpha_{\rm stat}/(V_{0}\eo)}{1-4\pi\alpha_{\rm stat}/(3V_{0}\eo)}\right)\:,\label{eq:eeff1}
\end{equation}
where $\alpha_{\rm stat}$ is obtained from $\alpha$ by neglecting the radiative correction term $\propto q^{3}$ in the Green function Eq.~\eqref{eq:G0}. Alternatively, Eq.~\eqref{eq:eeff1} can be presented as
\begin{equation}
\frac{\eps_{\rm eff}-\eo}{\eps_{\rm eff}+2\eo}=v\frac{\eps_{\rm Si}-\eo}{\eps_{\rm Si}+2\eo}\:,
\end{equation}
where $v\equiv V_{\rm NC}/V_{0}$ is the volume fill factor of the nanocrystals.
Expressing the polarizability $\alpha_{\rm stat}$ via $\eps_{\rm eff}$ from Eq.~\eqref{eq:eeff1}, substituting it into Eq.~\eqref{eq:fgen} and calculating the inverse matrix, we obtain
\begin{multline}
f_{\rm purc}=\frac{3}{2(\omega/c)^{3}\alpha_{\rm bare}^{2}}\Im \sum\limits_{\bm k}
\frac{V(\ee-\eo)}{4\pi\ee}\\\times\frac{\ee (k_{x}^{2}+k_{y}^{2})+\eo k_{z}^{2}-(\omega/c)^{2}\ee\eo}{k^{2}-\ee (\omega/c)^{2}-\rmi 0}\:.
\end{multline}
The integration can be performed by introducing the spherical coordinate system for the wave vector $\bm k$.
The imaginary part is determined by the residue at $k=\omega\sqrt{\ee}/c$. The Purcell factor can be presented as
\begin{align}
f_{\rm purc}&=\sqrt{\ee}F(\eout)\:,\label{eq:final1}\\
\frac{1}{\eout}&=\frac{v}{\ei}+\frac{1-v}{\eo}\label{eq:final2}\:.
\end{align}
Equations~\eqref{eq:final1},\eqref{eq:final2} constitute the central result of this study.
\section{Results and discussion}\label{sec:Discussion}
\begin{figure}[t]
\includegraphics[width=0.95\columnwidth]{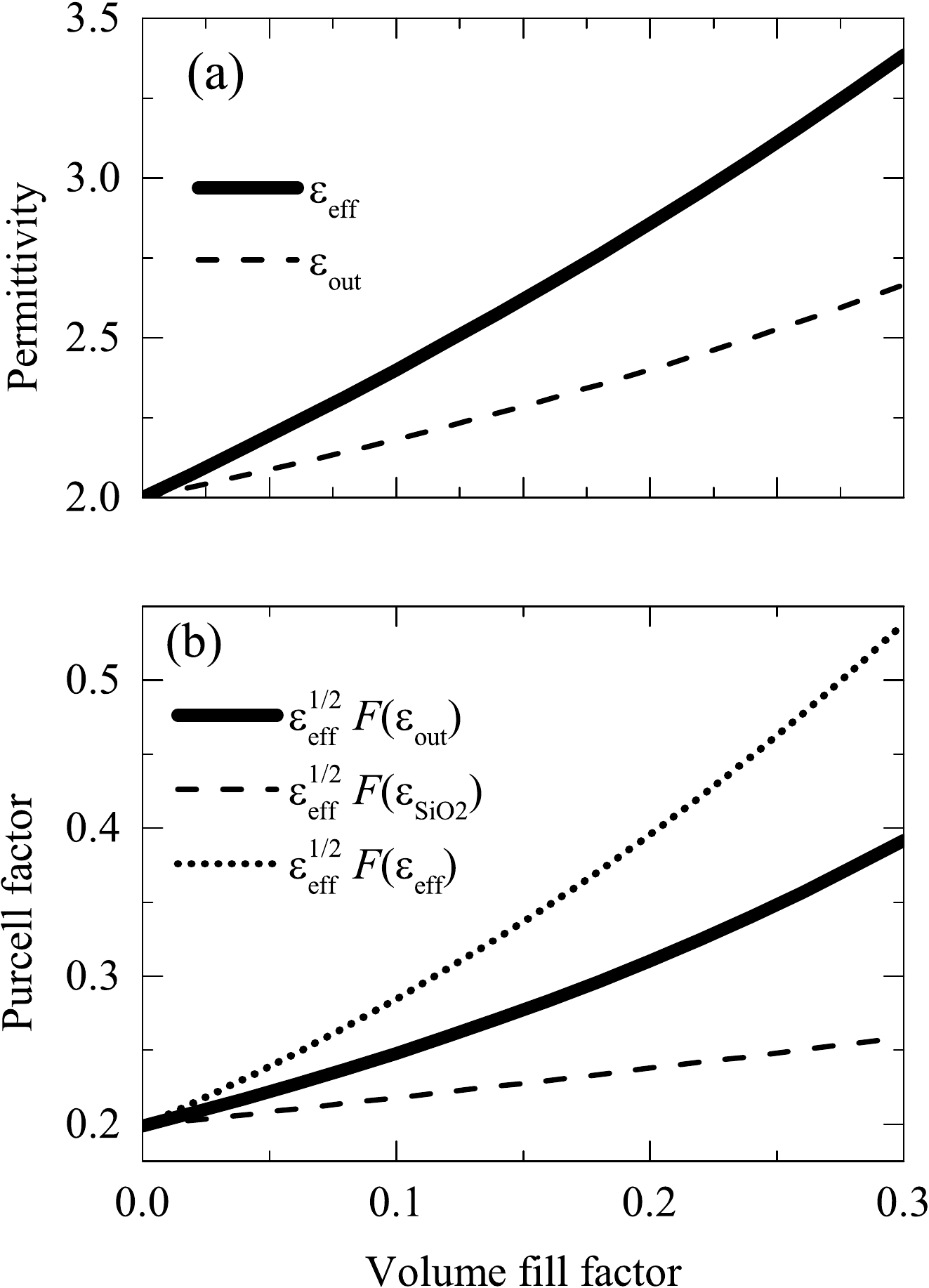}
\caption{(a) Dependence of the effective permittivities $\ee$ Eq.~\eqref{eq:eeff1} and $\eout$ Eq.~\eqref{eq:final2} on the volume fill factor of the nanocrystals $v$.
(b) Dependence of Purcell factor $f_{\rm purc}$ on the volume fill factor. 
Solid, dashed and dash-dotted curves correspond to  $\sqrt{\ee}F(\eout)$ (rigorous answer), and the naive approximations $\sqrt{\ee}F(\eo)$ and $\sqrt{\ee}F(\ee)$, respectively.
Calculation has been performed for $\ei=12$ and $\eo=2$.
}\label{fig:2}
\end{figure}

We proceed to the detailed analysis of the Purcell factor for the array of nanocrystals Eq.~\eqref{eq:final1}. For vanishing fill factor of the nanocrystals, $v\to 0$, Eq.~\eqref{eq:final1} reduces to the result for a single nanocrystal, Eq.~\eqref{eq:rad1}, because $\eout=\ee=\eo$. In the general case Eq.~\eqref{eq:rad1} and
Eq.~\eqref{eq:final1} have two important differences. First, the factor $\sqrt{\eo}$ in Eq.~\eqref{eq:rad1} is replaced by $\sqrt{\ee}$ in  Eq.~\eqref{eq:final1}. This reflects the modification of the density of photonic states in the medium. Instead of the dielectric constant of the empty matrix $\eo$, the structure is described by the effective dielectric constant of the nanocrystal array $\ee$. Second, the local-field factor $F(\eo)$ in Eq.~\eqref{eq:rad1}  is replaced by the factor $F(\eout)$ in Eq.~\eqref{eq:final1} with the permittivity $\eout$ given by the harmonic mean Eq.~\eqref{eq:final2}. 
Since $\ei> \eo$, the considered permittivities satisfy the following inequality,
\begin{equation}
\eo<\eout<\ee<\ei \label{eq:epsilons}
\end{equation}
which means that
\begin{equation}
 F(\eo)<F(\eout)<1\:.\label{eq:tF}
\end{equation}
Hence, the role of local field corrections is suppressed for the array of nanocrystals as compared to the empty matrix. This result is quite natural: since the local permittivity outside the nanocrystal increases, the dielectric contrast between the nanocrystal and its surroundings is reduced, and hence, the factor $F$ becomes closer to unity. Counterintuitively, the  harmonic mean permittivity $\eout$, that determines the local field corrections $F(\eout)$, is different from the effective permittivity $\ee$, that determines the density of photonic states. This can be related to the difference between the far field (described by $\ee$) and the near field (described by $\eout$). The far field probes the modification of the dielectric environment at the spatial scale larger than the light wavelength, while the near field is sensitive to the immediate environment of the emitting nanocrystal. 

These results are illustrated in Fig.~\ref{fig:2}. Panel (a) presents the dependence of the effective dielectric constants $\ee$ (solid curve) and $\eout$ (dashed curve)  on the volume fill factor $v$.
Both permittivities grow with $v$ from $\eo$ towards $\ei$, however, $\eout$ is noticeably smaller. 
It should be noted that both the Maxwell-Garnett approximation and the coupled-dipole model are not applicable
for large densities of nanocrystals when the deviation from the point dipole approximation become important~\cite{Milton}. Hence, we limit the range of fill factors in Fig.~\ref{fig:2} by the value $v=0.3$.  Figure~\ref{fig:2}(b) shows the dependence of the Purcell factor on $v$. The values are smaller than unity due to the strong screening of the field, described by the factor $F$. The solid curve has been calculated according to Eq.~\eqref{eq:final1}. The dashed curve has been calculated as $f_{\rm purc}=\sqrt{\ee}F(\eo)$, i.e. neglecting the suppression of the local field corrections for large fill factor and overestimating their importance. The dotted curve corresponds to $f_{\rm purc}=\sqrt{\ee}F(\ee)$, i.e. it underestimates the role of the local field corrections. We see, that the three curves considerably differ for $v\sim 0.2$.
Hence, the modification of the local field  corrections in the nanocrystal array is an important effect for such relatively dense nanocrystal samples.
 Our rigorous result Eq.~\eqref{eq:final1} corresponds to the   value of the Purcell factor, intermediate between the two naive approximations.

\section{Summary}\label{sec:Summary}
We have developed a theory of spontaneous emission in the dense arrays of Si nanocrystals. By using the discrete dipole approximation we were able to obtaine a closed form analytical answer for the Purcell factor, valid in the long-wavelength approximation when both the size of the nanocrystals and the spacing between them are much smaller than the wavelength of light. This expression reflects  (i) the enhancement of  the density of photonic states and (ii) the suppression of the local-field corrections when the volume fill factor of the nanocrystals in the sample increases. We demonstrate, that these two effects are described by two different effective permittivities, namely, the Maxwell-Garnett permittivity and the harmonic mean permittivity. While the model is relatively crude and does not account for the disorder, inevitably present in the actual samples, it still provides an important insight in the problem.
Our results could be instrumental in optical characterization of the emission kinetics of nanocrystal sample. Moreover, the theory is quite general and could be applied, for instance, for the rapidly developing area of all-dielectric silicon-based nanophotonics~\cite{Kuznetsov2012,Staude2013}.

\section*{Acknowledgements}
This work has been supported by the  Russian Science Foundation grant No. 14-12-01067.
I acknowledge useful discussions with R.~Limpens.


\end{document}